\documentclass{svmult} 

\usepackage{graphicx}        

\begin{document} 

\title*{ A discrete curvature on a planar graph}
\author{Miguel Lorente}
\institute{Departamento de F\'{\i}sica, Universidad de Oviedo, 33007 Oviedo, Spain
\texttt{lorentemiguel@uniovi.es}}
\maketitle
\begin{center}
{\it \small Dedicated to Alberto Galindo on the occasion of his 70th birthday. November 2004}
\end{center}

\begin{abstract} 
Given a planar graph derived from a spherical, euclidean or hyperbolic tessellation, one can define a discrete curvature by combinatorial properties, which after
embedding the graph in a compact 2d-manifold, becomes the Gaussian curvature.
\end{abstract}

\section{Introduction}

In recent years some approaches to quantum gravity have suggested the hypothesis of a discrete space time [1] as a consequence of the combinatorial properties of
spin networks underlying the structure of space [2] and implemented with the discrete time hypothesis of causal sets [3]. We have also presented a philosophical
interpretation of a discrete model of space time, based in the relational theory proposed by Leibniz [4].

In this model one starts from a set of elements and the relation among them, without presupposing space and time as a background. We have only at our disposal
combinatorial properties of those relations, from which the metrical quantities should be generated.

As Riemann mentioned in his Habilitation Thesis: ``The quantitative comparison [of quanta of space] happens for discrete magnitudes through counting, for
continuous ones through measurements'' [5].

In our model we have to choose the discrete quantities in such a way that in the continuous limit they become the classical ones. To this aim we present a two ways
method. The direct way consists on calculating some continuous quantity in a 2d-manifold. For instance, take a tessellation generated by some triangle reflection
group. Keeping only the vertices and edges of the tessellation, we obtain a graph (the skeleton) where we can define discrete quantities such as path, distance,
curvature. 

By the inverse way, we construct an embedding of the graph such that we recover the continuous quantities. In this example, the embedding surface is the original
manifold of the tessellation. We can start directly from the graph and find some embedding where the corresponding quantities become analog, like the genus of some
3nj-symbols graph [6], or the curvature in a triangulated manifold [7] or the area and volume eigenvalues for the gravitational field operator [8]. According
to Bombelli one can scatter points in a Lorentzian manifold with uniform density, and then keep the statistical distribution of points from which some discrete
quantities can be defined, such as curvature, from combinatorial properties of the set of relations [9].

In section 2 we review algebraic properties of triangle reflection groups in 2d-manifolds. In section 3 we present the spherical, euclidean hyperbolic
tessellations generated by Coxeter groups. In section 4 we apply the fundamental properties of Gauss curvature to the continuous tessellations. In section 5 we
construct a graph and define on it the curvature such that, when the embedding is performed, we recover the standard one.

These ideas were communicated in the Symposium held in Marseille on loops and spin foams [10].

\section{Reflection groups}

Let $P$ a finite sided n-dimensional convex polyhedron in a metric space $X$ of finite volume, all of whose dihedral angles are submultiple of $\pi$. Then the
group generated by the reflection of $X$ in the sides of $P$ is a discrete reflection group $\Gamma$ with respect to the polyhedron $P$.

In order to construct a presentation for a discrete reflection group, we take all the sides $\left\{ {S_i} \right\}$ of $P$ for each par of indices $i, j$. Let
$k_{ij}\equiv {\pi  \over {\vartheta \left( {S_i,S_j} \right)}}$, where ${\vartheta \left( {S_i,S_j} \right)}$ is the angle between $S_i$ and $S_j$. We call
\begin{equation}
\left\{ {S_i,\left( {S_iS_j} \right)^{k_{ij}}} \right\}
\end{equation}
a presentation of the discrete reflection groups $\Gamma$ generated by the reflections on $S_i$.

A discrete reflection group is isomorphic to a Coxeter group $G$, that is, an abstract group defined by a group presentation $\left\{ {S_i,\left( {S_iS_j}
\right)^{k_{ij}}} \right\}$ where 

\begin{enumerate}
\item[a)]  the exponent $k_{ij}$  is a positive integer or infinite,
\item[b)] $k_{ij}=k_{ji}>1,$ if $i\ne j,\;k_{ii}=1$ for each $i,j$,
\item[c)] If $k_{ij}=\infty $, the term $\left( {S_iS_j} \right)^\infty $ is omitted
\end{enumerate}

The Coxeter graph of $G$ is the labeled graph with vertices $i\in J$ an edges $\left\{ {\left( {i,j} \right),k_{ij}>2} \right\}$. Each edge ${i,j}$ is labelled
by $k_{ij}$. For simplicity the label $k_{ij}=3$ is omitted.

Let $\Delta$ be an n-simplex in $X$ all of whose dihedral angles are submultiple of $\pi$. The group $\Gamma$ generated by the reflections of X in the sides of
$\Delta$ is an n-simplex discrete reflection group. Notice that $X$ can be $S^n,E^n\ {\rm or}\  H^n$.

The classification of all the irreducible n-simplex (spherical, euclidean and hyperbolic) reflection groups is complete [11].

Assume that $n=2$. Then $\Delta$ is a triangle in $X$, whose angles ${\pi  \over l},{\pi  \over m},{\pi  \over n}$ are submultiple of $\pi$. If we call $T\left(
{l,m,n} \right)$ the group $\Gamma$ generated by the reflections in the sides of $T\left(
{l,m,n} \right)$ is call a triangle reflection group. We considere all the cases:

If $X=S^2$ the only spherical triangle reflection groups have the following Coxeter graphs:

\setlength{\unitlength}{0.7mm}
\hspace*{.1cm}
$\underbrace{
\begin{picture}(30,10)
\put(0,2){\circle*{1.75}}
\put(15,2){\circle*{1.75}}
\put(30,2){\circle*{1.75}}
\end{picture}}_{T(2,2,2)}$
\hspace{2cm}
$\underbrace{
\begin{picture}(30,10)
\put(0,2){\circle*{1.75}}
\put(15,2){\circle*{1.75}}
\put(15,2){\line(1,0){15}}
\put(30,2){\circle*{1.75}}
\put(22,4){\scriptsize $n$}
\end{picture}}_{T(2,2,n)\ n>2}$
\hspace{2cm}
\begin{picture}(30,10)
\put(0,2){\circle*{1.75}}
\put(0,2){\line(1,0){15}}
\put(15,2){\circle*{1.75}}
\put(15,2){\line(1,0){15}}
\put(30,2){\circle*{1.75}}
\put(15,-4.7){\makebox(0,0){\scriptsize $T$(2,3,3)}}
\end{picture}

\bigskip 

\hspace*{.1cm}
$\underbrace{
\begin{picture}(30,2)
\put(0,2){\circle*{1.75}}
\put(0,2){\line(1,0){30}}
\put(15,2){\circle*{1.75}}
\put(30,2){\circle*{1.75}}
\put(22,4){\scriptsize $4$}
\end{picture}}_{T(2,3,4)}$
\hspace{2cm}
$\underbrace{
\begin{picture}(30,2)
\put(0,2){\circle*{1.75}}
\put(0,2){\line(1,0){30}}
\put(15,2){\circle*{1.75}}
\put(15,2){\line(1,0){15}}
\put(30,2){\circle*{1.75}}
\put(22,4){\scriptsize $5$}
\end{picture}}_{T(2,3,5)}$

\bigskip 

If $X=E^2$ we have the euclidean triangle reflection groups with Coxeter graphs:

\hspace*{.85cm}
$\underbrace{
\begin{picture}(15,20)
\put(0,2){\circle*{1.75}}
\put(7.5,14.5){\circle*{1.75}}
\put(15,2){\circle*{1.75}}
\put(0,2){\line(1,0){15}}
\put(0,2){\line(3,5){7.3}}
\put(15,2){\line(-3,5){7.3}}
\end{picture}}_{T(3,3,3)}$
\hspace{2,75cm}
$\underbrace{
\begin{picture}(30,2)
\put(0,2){\circle*{1.75}}
\put(15,2){\circle*{1.75}}
\put(0,2){\line(1,0){30}}
\put(30,2){\circle*{1.75}}
\put(7,4){\scriptsize $4$}
\put(22,4){\scriptsize $4$}
\end{picture}}_{T(2,4,4)}$
\hspace{2cm}
$\underbrace{
\begin{picture}(30,2)
\put(0,2){\circle*{1.75}}
\put(0,2){\line(1,0){15}}
\put(15,2){\circle*{1.75}}
\put(15,2){\line(1,0){15}}
\put(30,2){\circle*{1.75}}
\put(22,4){\scriptsize $6$}
\end{picture}}_{T(2,3,6)}$

\bigskip
If $X=H^2$ we have the hyperbolic triangle reflection groups with Coxeter graphs

\begin{center}
$\underbrace{
\begin{picture}(30,2)
\put(0,2){\circle*{1.75}}
\put(15,2){\circle*{1.75}}
\put(0,2){\line(1,0){30}}
\put(30,2){\circle*{1.75}}
\put(7,4){\scriptsize $m$}
\put(22,4){\scriptsize $n$}
\end{picture}}_{T(2,m,n)\ m\geq n\geq 3}$
\hspace{2,75cm}
$\underbrace{
\begin{picture}(15,15)
\put(0,2){\circle*{1.75}}
\put(7.5,14.5){\circle*{1.75}}
\put(15,2){\circle*{1.75}}
\put(0,2){\line(1,0){15}}
\put(0,2){\line(3,5){7.3}}
\put(15,2){\line(-3,5){7.3}}
\put(1,8){\scriptsize $l$}
\put(7,4){\scriptsize $n$}
\put(13,8){\scriptsize $m$}
\end{picture}}_{T(l,m,n)\ l \geq m \geq n \geq 3}$
\end{center}

Another type of Coxeter groups are generalized (or non-compact) simple reflection groups, that are defined only in $H^n$. A generalized n-simplex is an
n-dimensional polyhedron with $n+1$ generalized vertices (either a vertex of P). the generalized (non compact) hyperbolic n-simple reflection groups exist only
for $n\le 10$ and have been classified completely [12].

The generalized hyperbolic triangle reflection groups have the following Coxeter graph:

\setlength{\unitlength}{0.8mm}
\begin{center}
\begin{picture}(15,15)
\put(0,2){\circle*{1.75}}
\put(7.5,14.5){\circle*{1.75}}
\put(15,2){\circle*{1.75}}
\put(0,2){\line(1,0){15}}
\put(0,2){\line(3,5){7.3}}
\put(15,2){\line(-3,5){7.3}}
\put(0,8){\scriptsize $\infty$}
\put(7,-1){\scriptsize $\infty$}
\put(13,8){\scriptsize $\infty$}
\end{picture}
\hspace{2cm}
\begin{picture}(15,15)
\put(0,2){\circle*{1.75}}
\put(7.5,14.5){\circle*{1.75}}
\put(15,2){\circle*{1.75}}
\put(0,2){\line(1,0){15}}
\put(0,2){\line(3,5){7.3}}
\put(15,2){\line(-3,5){7.3}}
\put(0,8){\scriptsize $\infty$}
\put(7,-1){\scriptsize $l$}
\put(13,8){\scriptsize $\infty$}
\end{picture}
\hspace{2cm}
\begin{picture}(15,15)
\put(0,2){\circle*{1.75}}
\put(7.5,14.5){\circle*{1.75}}
\put(15,2){\circle*{1.75}}
\put(0,2){\line(1,0){15}}
\put(0,2){\line(3,5){7.3}}
\put(15,2){\line(-3,5){7.3}}
\put(0,8){\scriptsize $l$}
\put(7,-1){\scriptsize $\infty$}
\put(13,8){\scriptsize $m$}
\end{picture}
\hspace{1.5cm}
$l \geq m \geq 3$
\end{center}

\ 

\begin{center}
\begin{picture}(30,2)
\put(0,2){\circle*{1.75}}
\put(15,2){\circle*{1.75}}
\put(0,2){\line(1,0){30}}
\put(30,2){\circle*{1.75}}
\put(7,4){\scriptsize $\infty$}
\put(22,4){\scriptsize $\infty$}
\end{picture}
\hspace{2,75cm}
\begin{picture}(30,2)
\put(0,2){\circle*{1.75}}
\put(15,2){\circle*{1.75}}
\put(0,2){\line(1,0){30}}
\put(30,2){\circle*{1.75}}
\put(7,4){\scriptsize $n$}
\put(22,4){\scriptsize $\infty$}
\end{picture}
\hspace{2cm}
$n \geq 3$
\end{center}

\section{Geometric representation of Coxeter group and tessellations}

We have seen in Section 2 that a Coxeter group is isomorphic to a discrete reflection group. Geometrically a reflection can be represented by a linear
transformation which fixes an hyperplane point wise and sends some non zero vector to its negative. In the metric space $X$ we construct vectors $\left\{
{\alpha _i} \right\}$ in one to one correspondence to the sides $\left\{ {S_i} \right\}$ defined before, in such a way that the angle between ${\alpha _i}$
and ${\alpha _j}$ will be compatible with the values of $k_{ij}$, namely, $\vartheta \left( {\alpha _i,\alpha _j} \right)={\pi  \over {k_{ij}}}$.

In order to construct a reflection with respect to these vectors $\left\{ {\alpha _i} \right\}$ we define a non-degenerate symmetric bilinear form on $X$ by
the formulas
\begin{equation}
\left\langle {\alpha _i,\alpha _j} \right\rangle =-\cos {\pi  \over {k_{ij}}}
\end{equation}
This expresion is interpreted to be $-1$ for $k_{ij}=\infty $. Obviously $\left\langle {\alpha _i,\alpha _i} \right\rangle =1$, and $\left\langle {\alpha
_i,\alpha _j} \right\rangle \le 0$ for $i\ne j$. For each vector ${\alpha _i}$ we can define a reflection ${S_i}$ on $X$:

$$S_i\beta =\beta -2\left\langle {\alpha _i,\beta } \right\rangle \alpha _i\quad ,\quad \beta \in X$$
clearly $S_i\alpha _i=-\alpha _i$ and all the vectors $\gamma $ satisfying $\left\langle {\alpha _i,\gamma } \right\rangle =0$ belong to a plane invariant
under $S_i$. It can be proved that the group of reflections generated by $\left\{ {S_i} \right\}$ is homomorphic to the Coxeter group $G$ because they satisfy 
\begin{equation}
S_i^2=1\;,\quad \left( {S_iS_j} \right)^{k_{ij}}=1\quad {\rm for}\;\,i\ne j
\end{equation}
we call this homomorphism the geometric representation of $G$.

The group of reflections generated by $\left\{ {S_i} \right\}$ is connected with the geometrical property of space called tessellation. This can be seen in an
intuitive way by the covering of the, say, two dimensional euclidean space, repeating the same or a finite number of figures without overlaping or holes.

In general, a tessellation of a metric space $X\left( {=S^n,E^n\,{\rm or}\,H^n} \right)$ is a collection $\mathcal{P}$ of n-dimensional convex polyhedra in $X$
such that.

\begin{enumerate}
\item[a)]  the interior of the polyhedra in $\mathcal{P}$ are mutually disjoint,
\item[b)] the union of the polyhedra in $\mathcal{P}$ is $X$.
\end{enumerate}

For our purpose we need two more definitions.

A tessellation of $X$ is exact if and only if each side of a polyhedron $P$ in $\mathcal{P}$ is a side of exactly two polyhedra $P$ and $Q$ in $\mathcal{P}$.

A regular tessellation of $X$ is an exact tessellation of $X$ consisting of congruent regular polytopes.

We have shown in section 2 that if $\Delta $ is an n-simplex in $X$ all of whose dihedral angles are submultiple as $\pi$ then the group  $\Gamma $ generated
by the reflection of $X$ in the sides of $\Delta $ is a discrete reflection group, the geometrical representation of which was given with the help of the
bilinear form (2).

It can be proved that the collection of the polyhedra obtained by the reflections on the side of $\Delta $ is a tessellation of $X$. Since an n-simplex is a
regular congruent polytope the tessellation is regular. Therefore all the n-simplex (compact or non-compact) reflection groups lead to regular tessellation of
$X$.

We give now some examples of tessellations in $S^2,E^2$ and $H^2$ generating by reflecting in the sides of a spherical, euclidean and hyperbolic triangle as
defined in section 2.

Notice that the hyperbolic tessellation has been drawn using te conformal disk model [13]

\begin{figure}
\centering
\includegraphics[height=4cm]{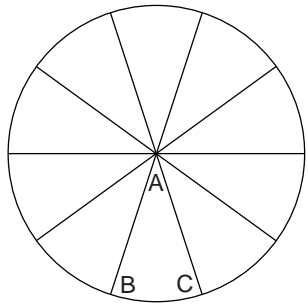}
\caption{Tessellation of $S^2$ (in stereographic projection) by reflecting in the sides of the spherical triangle $T(2,2,5)$}
\label{fig:4}       
\end{figure}
\begin{figure}
\centering
\includegraphics[height=5cm]{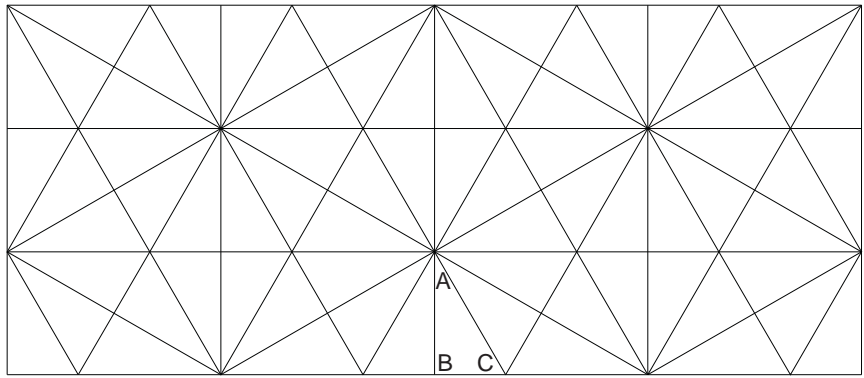}
\caption{Tessellation of $E^2$ generated by reflecting in the sides of the euclidean triangle $T(2,3,6)$}
\label{fig:4}       
\end{figure}

\begin{figure}
\centering
\includegraphics[height=8cm]{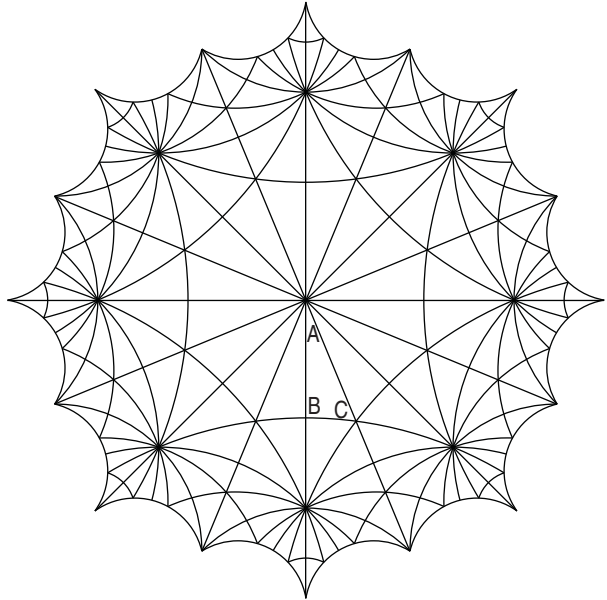}
\caption{Tessellation of $H^2$ generated by reflecting in the sides of the hyperbolic triangle $T(2,3,8)$}
\label{fig:4}       
\end{figure}

\section{Gauss curvature of continuous tessellations}

Two dimensional tessellations in $X\left( {=S^2,E^2\,{\rm or}\,H^2} \right)$ are generated by 2-simplex (triangle) reflection group. In order to calculate the
gaussian curvature, we review some geometrical properties of geodesic triangles.

In $S^2$ the geodesic triangle (that is, triangle whose sides are arcs of geodesics) are spherical. Given a spherical triangle $T\left( {x,y,z} \right)$ with:

\begin{center}
\includegraphics[height=4cm]{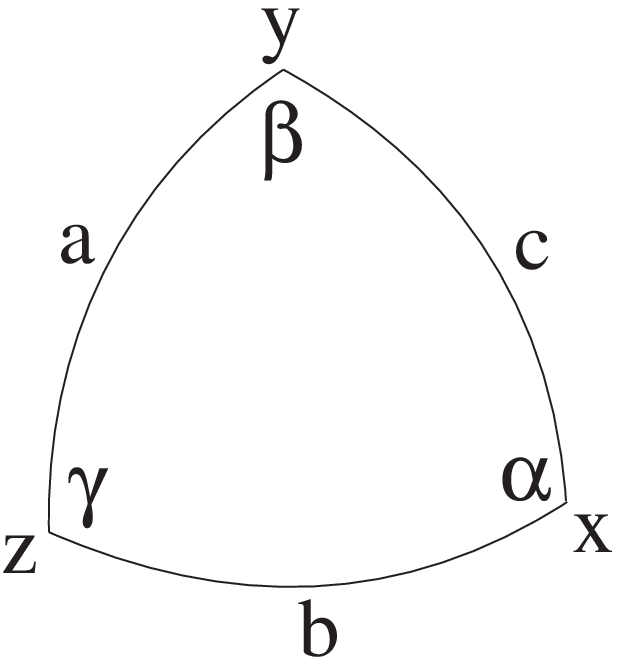}
\end{center}
$\begin{array}{llll}
\mbox{Geodesic sides:} &\left[ {y,z} \right], &\left[ {z,x} \right], &\left[ {x,y} \right],\\
\mbox{Length of sides:} & a\;, & b\;,& c\;, \\
\mbox{Geodesic arcs:} & f(t) & g(t) & h(t) \\
 & \left[ {o,a} \right]\to S^2, & \left[ {o,b} \right]\to S^2, & \left[ {o,c} \right]\to S^2
\end{array}$

Angle ${\alpha }$ between $\left[ {z,x} \right]$ and $\left[ {x,y} \right]=\vartheta \left( {-g'(b),h'(0)} \right)$

Angle ${\beta }$ between $\left[ {x,y} \right]$ and $\left[ {y,z} \right]=\vartheta \left( {-h'(c),f'(0)} \right)$

Angle ${\gamma }$ between $\left[ {y,z} \right]$ and $\left[ {z,x} \right]=\vartheta \left( {f'(a),g'(0)} \right)$

The excess of the interior angles of a spherical triangle is:
\begin{equation}
\in =\alpha +\beta +\gamma -\pi 
\end{equation}
It can be proved that this excess is always positive [14].

The area of the triangle $T\left( {x,y,z} \right)$ is [15]
\begin{equation} 
{\rm Area} \left\{ {T\left( {x,y,z} \right)} \right\}=\alpha +\beta +\gamma -\pi =\in
\end{equation}
In $E^2$ we have euclidean triangles $T\left( {x,y,z} \right)$ 

\begin{center}
\includegraphics[height=3cm]{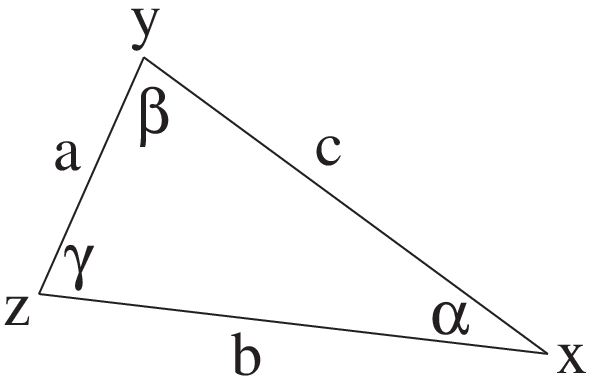}
\end{center}

$\begin{array}{llll}
\mbox{with sides:} &\left[ {y,z} \right], &\left[ {z,x} \right], &\left[ {x,y} \right],\\
\mbox{Length of sides:} & a\;, & b\;,& c\;, \\
\mbox{Angles:} & \alpha , & \beta , & \gamma  
\end{array}$

The excess of the interior angles of an euclidean triangle is 
\begin{equation} 
\in =\alpha +\beta +\gamma -\pi =0
\end{equation}
In $H^2$ the geodesic triangles are hyperbolic. For the hyperbolic triangle  $T\left( {x,y,z} \right)$ we have 

\begin{center}
\includegraphics[height=3cm]{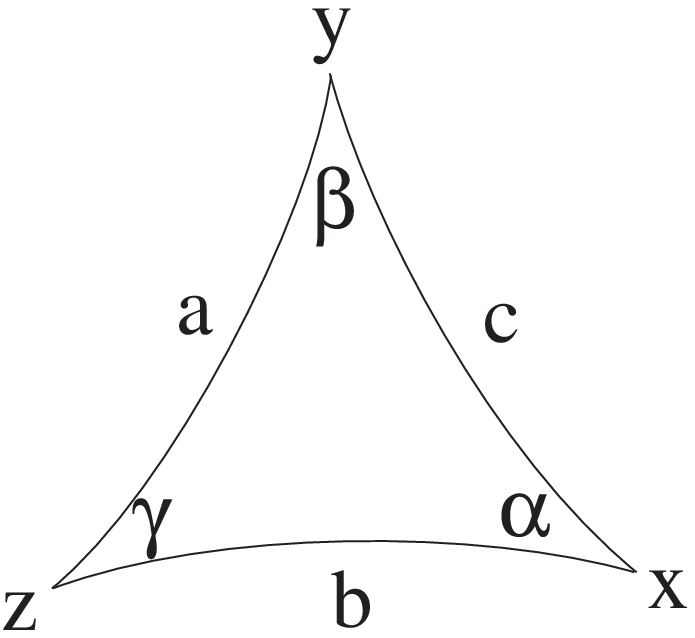}
\end{center}

$\begin{array}{llll}
\mbox{Geodesic sides:} &\left[ {y,z} \right]\;, &\left[ {z,x} \right]\;, &\left[ {x,y} \right]\;, \\
\mbox{Length of sides:} &a\;, &b\;, &c\;, \\
\mbox{Geodesic arcs:} &f(t)\;, &g(t)\;, &h(t)\;, \\
 &\left[ {o,a} \right]\to H^2, &\left[ {o,b} \right]\to H^2, &\left[ {o,c} \right]\to H^2
\end{array}$

Angle between $\left[ {z,x} \right]$ and $\left[ {x,y} \right]: \alpha=\vartheta \left( {-g'(b),h'(0)} \right)$

Angle between $\left[ {x,y} \right]$ and $\left[ {y,z} \right]: \beta=\vartheta \left( {-h'(c),f'(0)} \right)$

Angle between $\left[ {y,z} \right]$ and $\left[ {z,x} \right]: \gamma=\vartheta \left( {-f'(a),g'(0)} \right)$

A generalized hyperbolic triangle has at least one vertex at infinite. The corresponding angle is zero. An ideal triangle has three vertices at infinite. The
excess of the interior angles of an hyperbolic triangle is
\begin{equation} 
\in =\alpha +\beta +\gamma -\pi
\end{equation}

It can be proved that this excess is always negative [16]

The area of an hyperbolic triangle $T\left( {x,y,z} \right)$ is [17]
\begin{equation} 
{\rm Area}\;(T)=\pi -\left( {\alpha +\beta +\gamma } \right)
\end{equation}

If $T\left( {x,y,z} \right)$ is an generalized hyperbolic triangle
\begin{equation} 
{\rm Area}\;(T)=\pi -\alpha
\end{equation}

If $T\left( {x,y,z} \right)$ is an ideal triangle
\begin{equation} 
{\rm Area}\;T\left( {x,y,z} \right)=\pi
\end{equation}

We now give explicit values for the excess angle and area of the 2-simplex (triangle) reflection groups.

In $S^2$ the spherical triangle are given in Section 2.

The corresponging excess and area are, in units of $\pi$ $\left( {\rm remember:}\alpha ={\pi  \over l},\beta ={\pi  \over m},\gamma ={\pi  \over n} \right):$
 
\begin{equation}
{\epsilon  \over \pi }={1 \over l}+{1 \over m}+{1 \over n}-1
\end{equation}
\begin{equation}
{A \over \pi }={1 \over l}+{1 \over m}+{1 \over n}-1
\end{equation}

We have the following cases

$$\begin{array}{ll}
T\left( {2,2,2} \right): &{\epsilon  \over \pi }={A \over \pi }={1 \over 2}\\[2mm]
T\left( {2,2,n} \right): &{\epsilon  \over \pi }={A \over \pi }={1 \over n}\\[2mm]
T\left( {2,3,3} \right): &{\epsilon  \over \pi }={A \over \pi }={1 \over 6}\\[2mm]
T\left( {2,3,4} \right): &{\epsilon  \over \pi }={A \over \pi }={1 \over {12}}\\[2mm]
T\left( {2,3,5} \right): &{\epsilon  \over \pi }={A \over \pi }={1 \over {30}}
\end{array}$$

In $E^2$ the euclidean triangles are given in Section 2.

The excess angle is $\epsilon=0$ for

$$T\left( {3,3,3} \right),\quad T\left( {2,4,4} \right)\quad {\rm and}\quad T\left( {2,3,6} \right)$$

In $H^2$ the hyperbolic triangles are given in Section 2.

The corresponding excess and area in units of $\pi$ are  $\left( {{\rm remember:}\alpha ={\pi  \over l},\beta ={\pi  \over m},\gamma ={\pi  \over n}} \right):$
\begin{equation}
{\epsilon  \over \pi }={1 \over l}+{1 \over m}+{1 \over n}-1
\end{equation}
\begin{equation}
{A \over \pi }=1-\left( {{1 \over l}+{1 \over m}+{1 \over n}} \right)
\end{equation}

When some of the vertices are ideal, the corresponding angle is zero (the number $l,m$ or $n$ becomes infinite).

For instance, $T\left( {l,m,\infty} \right),$
\begin{equation}
{\epsilon  \over \pi }={1 \over l}+{1 \over m}-1
\end{equation}
\begin{equation}
{A \over \pi }=1-\left( {{1 \over l}+{1 \over m}} \right)
\end{equation}

We can now apply these results to the curvature of the surfaces corresponding to the 2-dimensional regular tessellations (spherical, euclidean or hyperbolic).
According to Gauss-Bonet theorem [18] the excess angle of some geodesic triangle $T$ is equal to the integral of the gaussian curvature over $T$
\begin{equation}
\epsilon =\alpha +\beta +\gamma -\pi =\int\!\!\!\int\limits_T {Kd\sigma }
\end{equation}
where $d\sigma$ is the area element. If $K$=const.
\begin{equation}
K={\epsilon  \over A}
\end{equation}

Applying this formula to the above results, we have:

$K=1$ for spherical geodesic triangles

$K=0$ for euclidean triangles

$K=-1$ for hyperbolic geodesic triangle

We can use another interpretation of Gaussian curvature in terms of parallel transport. If $\Delta \varphi $ is the change of angle in the paraller transport
of a vector along a curve $C$ the trace of which is the boundary of a region $R$, containing the point $p$, then 
\begin{equation}
\Delta \varphi =\int\!\!\!\int\limits_R {Kd\sigma }
\end{equation}

Since $\Delta \varphi $ does not depend on the choice of $C$ (but it depends on the enclosed area $A(R)$)
\begin{equation}
\mathop {\lim }\limits_{R\to p }{{\Delta \varphi } \over {A\left( R \right)}}=K\left( p  \right)
\end{equation}

This formula gives a method to calculate the curvature at a point in terms of the area and the parallel transport along the border.

\section{Curvature on planar graphs}

A graph is a par $G=\left\{ {V,E} \right\}$ where $V$ is a non-empty set of vertices and $E$ an unordered 2-set of vertices, called edges, in such a way that
two vertices are incident to an edge. We exclude loops, which are edges incident twice with the same vertex, and parallel edges, which are pair of edges
incident with the same vertices. We are interested in planar sets, that is to say, a graph that can be drawn on a piece of paper, such that its edges intersect
only at their common vertices.

A graph can be defined in an abstract way using only combinatorial properties of vertices and edges, or can be obtained from geometrical objects. For instance,
given a particulr tessellation described in section 3, we keep the edges and vertices of all the triangles and eliminate the embedding manifold (in our case the
surface $S^2$, $E^2$ or $H^2$) in such a way that we are left with the points (vertices) and relations among than (edges). In Figures 4, 5, 6, we have drawn
the graphs that we have derived by this method from the tessellations given in Figures 1, 2, 3 respectively, where the vertices are represented by points and
the edges by arrows.

\begin{figure}
\centering
\includegraphics[height=4cm]{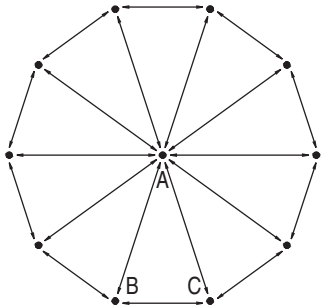}
\caption{Graph obtained from spherical tessellation of Fig. 1.}
\label{fig:4}       
\end{figure}

\begin{figure}
\centering
\includegraphics[height=5cm]{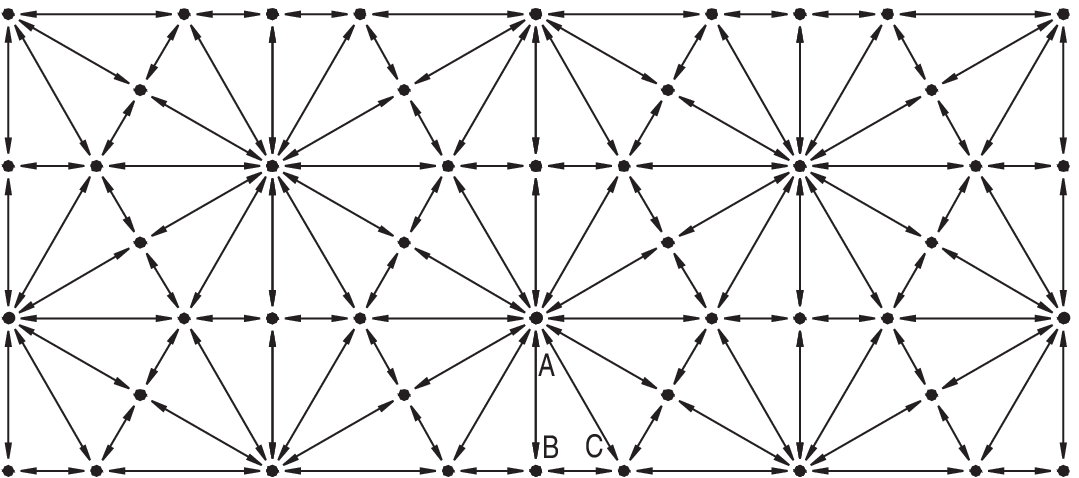}
\caption{Graph obtained from hyperbolic tessellation of Fig. 2.}
\label{fig:5}       
\end{figure}

\begin{figure}
\centering
\includegraphics[height=10cm]{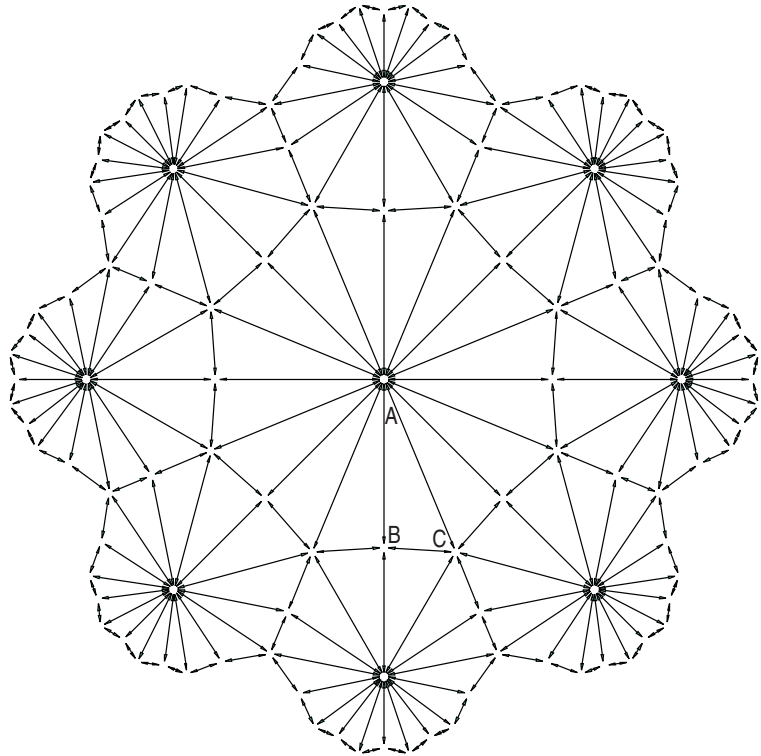}
\caption{Graph obtained from hyperbolic tessellation of Fig. 3.}
\label{fig:6}       
\end{figure}

In a graph one can define such elements as path, circuit, length, distance and other operations on graphs in analogy to the continuous case. For instance, in a
given graph one may travel from one vertex to another using several edges. the set of the vertices visited in that journey is called a path. The distance
between two vertices of a graph is the length of the shortest path between those two vertices. These definitions coincide with the standard ones when the graph
is embedded in some continuous manifold.

Given a planar graph corresponding to Fig. 4, 5, 6 where two adjacent vertices have always the same adjacent third vertex (different from the first two) one
can define the excess of this triad of vertices as the quantity, in analogy with (11), 
$$\delta ={1 \over l}+{1 \over m}+{1 \over n}-1$$
where $2l, 2m, 2n$ are the number of edges incident in each of the three vertices, which correspond to $2l$--valued, $2m$--valued or $2n$--valued vertices,
respectively. For instance, in Fig. 4, $A,B,C$ are 10--valued, 4--valued, 4--valued vertices, respectively; in Fig. 5, $A,B,C$ are 12--valued, 4--valued,
6--valued vertices, respectively; and in Fig. 6, $A,B,C$ are 16--valued, 4--valued, 6--valued vertices, respectively.

If we define the spherical, euclidean or hyperbolic graph, that is obtained from a spherical, euclidean or hyperbolic tessellation respectively, we can check

$\delta >0$, for a spherical graph (Fig. 4)

$\delta =0$, for an euclidean graph (Fig. 5)

$\delta <0$, for an hyperbolic graph (Fig. 6)

In a similar way, we can define the areas and curvature of a triad in a graph, such that they become the standard quantities when the graph is embedded in a
continuous manifold.

Therefore, we define the area of the triad $T(l, m, n)$ in a spherical graph, in analogy with (12),
$$\sigma (T)={1 \over l}+{1 \over m}+{1 \over n}-1$$
and the area of the triad $T(l, m, n)$ in an hyperboolic graph 
$$\sigma (T)=1-\left( {{1 \over l}+{1 \over m}+{1 \over n}} \right)$$

Similarly, we define the curvature of a triad $T(l, m, n)$

$$K(T)={\delta  \over \sigma }=\left\{ \begin{array}{lr}
{1,} &{\rm for\ a \ spherical\ graph} \\
{0,} &{\rm for\ an\ euclidean\ graph} \\
{-1,} &{\rm for\ an\ hyperbolic\ graph}
\end{array} \right.$$
an expression that can be considered the discrete version of the Gauss-Bonet theorem (17). As in the continuous case, the curvature of a graph at a vertex,
can be calculated as the parallel transport of a path surrounding the m-valued vertex divided by the area of the triads embraced by the path see (20).
Obviolusly 

$$K(P)={{2m\delta } \over {2mA}}=K(T)$$

\section{Some comments}
The method outline above can be applied to other more complicated graphs. For ins\-tance, in a non regular graph (not derived from a regular tessellation) some
discrete Gaussian curvature can be defined in terms of the $l, m$ or $n$-valued vertices of each triad, that leads to positive, zero or negative curvature for that
triad.

Also a more difficult task would be to apply our method to a non-planar graph coming from an n-dimensional tessellation generated by an n-simplex reflection
groups. As in the continuous case one can calculate the {\it sectional} curvature, that is obtained from some 2-dimensional surface which is the intersection of a
the space $X (S^n, E^n
\mbox or H^n)$ with the hyperplane perpendicular to the Weyl vector that generates a Coxeter reflection.

\section{Acknowledgments}
The author want to express his gratitude to the organizers Professors Rovelli and Freidel for the invitation to attend the Symposium held in May 2004 in
Marseille on ``Loops and spin foams in Quantum Gravity'', where he had the priviledge to present his ideas to Professor Smolin, Barrett, Bombelli, Gambini,
Markopoulou, Pullin, Sudarsky.

This work was partially supported by Ministerio Educaci\'on y Ciencia, grant BFM 2003-00313/FIS.

\end{document}